\documentclass[noeprint,noshowpacs,nopreprintnumbers,twocolumn,prb,showpacs,%
amsmath,amssymb,citeautoscript,aps,10pt]{revtex4-1}
\usepackage{graphicx}
\usepackage{dcolumn}
\usepackage{color}
\usepackage{bm}
\usepackage[hidelinks]{hyperref}
\hypersetup{
    colorlinks,
    citecolor=blue,
    filecolor=blue,
    linkcolor=blue,
    urlcolor=blue
}

\newcommand{\im}{\mathrm{i}}% Imaginary number "i"
\newcommand{\e}{\textrm{e}}% Natural number "e"

\DeclareMathOperator{\Tr}{Tr}

\allowdisplaybreaks

\usepackage{bm,dsfont}
\usepackage[integrals]{wasysym}
\usepackage{physics}
\newcommand\pF{p_{\mathrm F}}
\newcommand\vF{v_{\mathrm F}}
\newcommand\eF{\varepsilon _{\mathrm F}}

\newcommand\LR[3]{\left#1{#2}\right#3}
\newcommand\diag{\operatorname{diag}}
\newcommand{\dphi}{\widetilde \phi}

\def\lr#1#2#3{\left#1{#3}\right#2}
\begin{document}

\title{One-Dimensional Transport of Ultracold Bosons}

\author{Igor V. Lerner}
\email{i.v.lerner@bham.ac.uk}
\affiliation{School of Physics \& Astronomy, University of Birmingham, B15 2TT, UK}

\date{\today}

\begin{abstract}
Advances in cooling and trapping of atoms have enabled unprecedented experimental control of many-body quantum systems. This led to the observation of numerous quantum phenomena, important for fundamental science, indispensable for high-precision simulations of condensed-matter systems and promising for technological applications. However, transport measurements in neutral quantum gases are still in their infancy in contrast to the central role they play in electronics.   In these lectures, after reviewing nascent experiments on quantum fermionic transport, I will focus on our theoretical prediction sand the possibility of experimental observations of qualitatively new phenomena in transport of ultracold bosons which do not have a direct counterpart in quantum electronic transport in condensed matter systems. The description of this transport is based on the Luttinger liquid (LL) theory. So in the first part of the lectures I will introduce  main concepts of the LL based on the functional bosonisation approach.
\end{abstract}

\maketitle
\section{Introduction.}
Transport measurements make the backbone of experimental research into electronic properties of condensed matter systems. An extremely high accuracy of recording electrical currents and the possibility to control conductance by applying an external magnetic field, changing the carrier densities by gate voltages, etc, provided condensed-matter experimentalists with an unparallel versatility of methods, with measuring of response functions or $I$-$V$ characteristics being amongst the most fruitful ones.

None of these methods are directly available to researchers working with electrically neutral ultra-cold atoms or molecules. Although there were quite a few papers published on quantum transport in ultracold media over the decades, see e.g.\ \cite{1997:ultracoldtransport,2002:ultracoldtransport,2004:ultracoldtransport,2012:ultracoldtransport}, notably on the Anderson localisation \cite{Aspect2007a,Aspect2008,Inguscio:2008} and many-body localisation \cite{Schreiber:2015,Vidmar:2015,Kondov:2015}, the breakthrough studies of traditional for the condensed-matter physics response functions started only recently. \cite{Brantut:2012,Brantut:2013,Brantut:2015,2015:NatPhysQuantTransp} For neutral atoms, it was response current with respect to a difference of chemical (rather than electrochemical) potentials between two fermionic reservoirs, and the fact that the first of these papers reporting the measurement of Ohm's law \cite{Brantut:2012} has been published in \emph{Science} underlines the nontriviality of transport experiments in ultracold matter.

Although these first measurements revealed effects known for non-interacting fermions, they have opened a path for investigating correlation effects in quantum transport which are expected to be most pronounced in 1D systems, both for fermions and for bosons. A system of interacting particles in 1D with a linearised dispersion law \footnote{Such a linearisation is justified for characteristics driven primarily by the low-lying excitations in the vicinity of the Fermi level, or above the Bose--Einstein condensate.} is known as the Luttinger liquid (LL), and the model describing the low-lying excitations in 1D is called the Tomonaga--Luttinger model.\cite{Tom:50,*Lutt:63,*Haldane:81,*Haldane:81a} The  correlation effects in the LL result, in particular,  in a drastic modification of tunnelling into a 1D channel, or a flow through 1D channel across a single  impurity, as has been shown in numerous theoretical  works \cite{KaneFis:92a,*KaneFis:92b,MatYueGlaz:93,*FurusakiNagaosa:93b,%
*FabrizioGogolin:95,FurMatv:02,*NazGlaz:03,*PolGorn:03,*LYY:08,*GB:10} and experimental \cite{Bockrath:99,*Bockrath:01,*Yao:99,Auslaender:02,*Kim:06,*Levy:06,*Levy:2012} studies of electronic quantum transport.

\begin{figure}
\begin{center}
\includegraphics[width=0.47\textwidth]{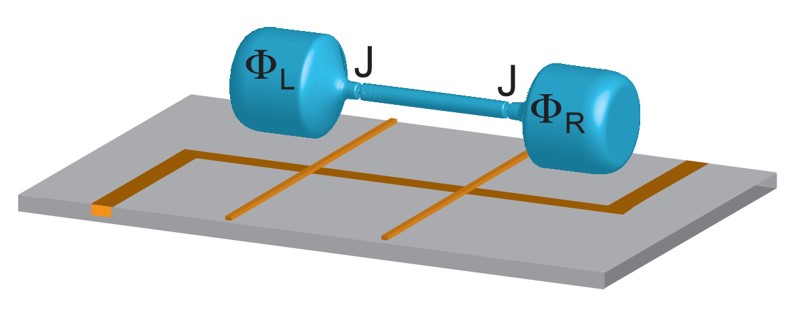}
\end{center}
\caption{\label{fig1}   A sketch of two 3D BEC reservoirs connected by a 1D channel via weak tunnelling links. The bosonic flow  is driven by the phase difference, $2\Phi =\Phi _{\mathrm{L}}-\Phi _{\mathrm{R}}$, between the reservoirs, while the chemical potentials a kept equal. A simplified illustration of an atom chip creating an appropriate magnetic trapping configuration is also shown.}
\end{figure}

In this lecture, I will present a theory \cite{SGKL} of transport of ultra-cold neutral bosons through a 1D channel connecting two BEC reservoirs via weak tunnelling links, whose role is similar to that of the impurity or a single weak link in the Luttinger liquid. The bosons are driven by a phase difference\footnote{The ways of creating such a phase difference are described in detail in Supplemental Online Material to the above cited paper.} between the reservoirs as schematically shown in Fig.~\ref{fig1}.

On the face of it, this problem has its condensed-matter counterpart: a  current between two bulk superconductors connected by a 1D channel via Josephson junctions    \cite{Fazio,*Fazio1,*MasStoneGoldbartLoss,*AffleckCauxZ}. However,   the bosonic flow drastically differs from  its electronic counterpart. In particular, the magnitude of the former is non-perturbative in the tunnelling strength $J$ in a sharp contrast to the latter. In many ways, in spite of the absence of any topological nontriviality, characteristics of the bosonic flow resemble those of the current through a 1D wire connecting two topologically nontrivial superconductors \cite{Lutchin:2010,*OregRefOpp} -- one of the firing ranges for studying Majorana fermions.

The cardinal difference between the 1D flow of interacting bosons and the 1D current of interacting electrons between topologically trivial superconductors is not due to the different particle exchange statistics but is owing to the difference in values of the so called Luttinger interaction parameter $K$: \ $K<1$ for  electrons with repulsion and $K\gg1 $ for   bosons with (a weak) repulsion. Hence, before describing the problem at hand in the second half of this lecture, in the first half I will derive  the Tomonaga--Luttinger model describing the low-lying excitations in 1D. The main approach to this is the bosonisation of such excitations; however, instead of the standard operator bosonisation  \cite{vDSh:98,Giamarchi} I will use an alternative -- the functional bosonisation. \cite{GYL:04}

\section{Functional bosonisation}
Linearisation of the standard quadratic dispersion law (as for electrons near the Fermi level) is useful in any dimensionality but paramount the  bosonisation of $1D$ systems.  Writing $$\mathcal{H}_0-\eF=\varepsilon ({p})-\eF=\frac{p^2-\pF^2}{2m} \approx \vF(|p|-\pF)\to \eta\vF q,$$ one sees that there are two almost independent energy branches in the vicinity of the Fermi level. They correspond two left- and right-movers, labelled by \mbox{$\eta=\pm \equiv  {\mathrm{R,L}}$,} with $q=\eta p-\pF$ counted from the appropriate Fermi point.

The linearised action ({with $\xi\equiv x,t$}) is then
 \begin{align}\notag
{\mathcal{S}}_0&=\int\!\mathcal{L}_0\dd\xi=\int\!\overline{\psi}({\xi })\LR [{\im \partial _t- ({{\mathcal{H}}-\eF})}]\psi({\xi})\dd x\\&=\mathcal{S}_{0+}+\mathcal{S}_{0-}\equiv \sum_{\eta=\pm}  \int\!\overline{\psi}_\eta({\xi }) \,\im \partial _\eta\psi_\eta({\xi})\dd x\,,   \label{S}
\end{align}
where the fields $\overline{\psi}$ and $\psi$ are matrix row and column, with $\overline{\psi}=\begin{pmatrix}
                   \overline{\psi}_+ & \overline{\psi}_- \\
                 \end{pmatrix}
$, and the linearised kernel of the action is the diagonal matrix, $\im\partial _t- ({{\mathcal{H}}-\eF})\to \diag ({\partial _+\,,\;\partial _-})$, with $\partial _\eta \equiv \partial _t-  \eta\vF\partial _x$. The correlation function of any quantum-mechanical observables is calculated via the functional integral, e.g.
\begin{align}\label{CF}
\begin{aligned}
    \langle{\mathcal{A}({\xi_1 })}\mathcal{A}({\xi _2})\rangle&=\frac{\int\,\mathcal{D}\overline{\psi}\mathcal{D}
\psi\,\mathcal{A}({\xi_1 })\mathcal{A}({\xi _2})\e ^{\im S_0}}{\int\,\mathcal{D}\overline{\psi}\mathcal{D}
\psi\,\e ^{\im S_0}}\,.
\end{aligned}
\end{align}
Effective excitations are made from (combinations of) left- and right- moving $e$-$h$ pairs which are bosonic.
Our aim is to show that the free particle Hamiltonian can be represented in terms of the bosonic fields $\theta_{\mathrm{L,R}}$ corresponding to the left and write movers as follows:
\begin{align}\label{H0}
    \mathcal{H}_0=\frac{\vF}{4\pi }\int\!\LR [{\LR ({\partial _x\theta_{\mathrm{L}}})^2+ \LR ({\partial _x\theta_{\mathrm{R}}})^2}]\dd x
\end{align}
where $\theta_{\mathrm{L,R}}$ are introduced via
 the following change of variables (``bosonisation''):
\begin{align}\label{change}
    \psi_\eta({\xi })=\chi _\eta{\mathrm{e}}^{\im \theta_\eta({\xi })}\,,\quad\overline{\psi} _\eta=\chi _\eta{\mathrm{e}}^{-\im \theta_\eta({\xi })}\,,
\end{align}
with $\chi _\eta$ being coordinate-independent Majorana fields, $\chi _\eta^2=0$. The idea of bosonisation is that all the dynamics of left- and right- movers is captured by the bosonic field $\theta_\eta({\xi })$, while the pre-exponential Majorana field (equivalent to the Klein factor in the operator bosonisation) is necessary to preserve the Fermionic statistics of the original fields $\psi_\eta({\xi })$ via the anticommutation relation \footnote{It is worth reminding that $\chi_\eta $ are fields rather than operators; for the latter, the anticommutation relation contains $\delta _{\eta,\eta'}$ in the right-hand side.}  $\{{\chi_\eta , \, \chi _{\eta'}}\}=0 $.

The ``bosonisation" of Eq.~\eqref{change} is useful even when the original fields $\psi_\eta$ are bosonic. In this case, its role is to represent the over-condensate bosonic excitations in term of slowly varying phase fields $\theta_\eta$.

However, on the face of it,
 substitution \eqref{change} leads,  in the case of fermions, to  the vanishing  action:
\begin{align}\label{L}
 \mathcal{S}_{0\eta} \to \widetilde{   \mathcal{S}}_{0\eta} =\int\!
\chi _\eta
\LR [{\im \partial _\eta-\alpha _\eta({\xi})}]
\chi _\eta \,\dd\xi =0,\,\;
\end{align}
where $\alpha _\eta\equiv \partial _\eta
\theta_\eta$. Indeed,  the first term vanishes since $\chi _\eta$ is coordinate-independent, and the second since $\chi _\eta^2=0$. This is the way in which the well-known anomaly in quantum field theory reveals itself in the functional formalism. It arises when dealing with the fields (or operators in the operator formalism) taken at the same point.

To shed light on the anomaly, let us consider the gauge transform,
 \begin{align}\label{gauge}
 \psi_\eta({\xi })=\widetilde{\psi}_\eta({\xi })\e ^{\im \theta_\eta({\xi })}  ,
\end{align}
where -- in contrast to Eq.~\eqref{change} -- $\theta_\eta({\xi })$ is an external field. Each part of the action,  $\mathcal{S}_0=\mathcal{S}_{0+}+\mathcal{S}_{0-}$, transforms similar to that in Eq.~\eqref{L}:
$$\int\overline{ {\psi}}_\eta({\xi })\LR [{\im \partial_{\eta} }] {\psi}_\eta({\xi })\dd\xi \to \int\overline{\widetilde{\psi}}_\eta({\xi })\LR [{\im \partial_{\eta}{-}\alpha _\eta({\xi })}]\widetilde{\psi}_\eta({\xi })\dd\xi,
$$
but it does not vanish.   The functional integral involves the Jacobian, $J(\alpha _\eta)$, of this transform that depends only the external field, $\alpha _\eta\equiv \partial_{\eta}\theta_\eta$, and can be found by calculating the partition function,
\begin{align}\label{J}
 \int\,\mathcal{D}\overline{\psi}_\eta\mathcal{D}
\psi_\eta\,\e ^{\im \mathcal{S}_{0\eta}}%=\det({\im \partial _\eta})
=   J({\alpha _\eta})\int\,\mathcal{D}\overline{\widetilde{\psi}}_\eta\mathcal{D}
\widetilde{\psi}_\eta\,\,\e ^{\im  \widetilde{\mathcal{S}}_{0\eta}}
%=\det({\im \partial _\eta-\alpha _\eta})\,.
%
\end{align}
Both functional  integrals above are Gaussian. If the fields $\overline{\psi}$ and $\psi$ are the conjugate bosonic fields, such a functional Gaussian integral would be equal to the inverse determinant of the  action kernel, which is a generalisation of the Gaussian integral over complex variables, $\int\frac{\dd z^\ast\!\dd z }{2\pi}\e ^{\im z^*\!Az} = A^{-1} $.  It turns out that the Gaussian integrals over anticommuting  fermionic (Grassmann) fields   are equal \cite{Berezin,Popov}  to the determinant of the  kernel. As the kernel of $\mathcal{S_{0\eta}}$ in the left-hand-side of Eq.~\eqref{J} is simply $\im \partial _\eta$, then the appropriate fermionic integral is equal to $\det \im \partial _\eta$, while the integral in the right-hand-side is equal to $\det \LR [{\im \partial _\eta-\alpha _\theta}]$.

The derivative $\partial _\eta\equiv \partial _t-\eta\vF\partial _x$ in the above expressions should be understood as the appropriate (inverse) Green's function, which by definition obeys the time-dependent Schr\"{o}dinger equation for the right- ($\eta=+1$) or left- ({$\eta=-1$}) movers:
\begin{subequations} {
\begin{align}\label{gf1}
    \im \partial _\eta g_\eta({\xi -\xi '})=  \delta ({t-t'})\delta ({x-x'})\,.
\end{align}
Its Fourier-transform  is given by
\begin{align}\label{gf2}
 g_\eta({p,\varepsilon})=\LR ({\varepsilon-\eta\vF p+
\im \delta \operatorname{sign}\varepsilon})^{-1}.
\end{align}
}\label{gf}\end{subequations}
Thus, $\im \partial _\eta$ can be substituted by $g_\eta^{-1} ({\xi })$, but using formal expressions like $\partial  _\eta^{-1} $ for $g_\eta$ will be convenient for algebraic manipulations.  This results in  the following Jacobian of the  gauge transform \eqref{gauge}:
\begin{align}\label{J}
 J(\alpha _\eta)&= \frac{\det \im \partial _\eta}{\det\LR ({\im \partial _\eta-\alpha _\eta})}=\frac{\det g_\eta^{-1} }{\det \LR ({g_\eta^{-1}-\alpha _\eta })}.
\end{align}
Using the well-known identity, $\ln \det A=\Tr \ln A  $, gives
\begin{align}
\notag
J(\alpha_\eta)&= \e ^{\Tr\ln g_\eta^{-1}-\Tr\ln\LR [{g_\eta^{-1}-\alpha _\eta }] }\\&=
\e ^{-\Tr\ln \ \LR  ({1-\alpha _\eta g_\eta} )}=\e ^{-\sum_{n=2} ^\infty\Tr\ \frac{\LR  ({\alpha _\eta g_\eta })^n}{n}}
    ,\label{J2}
\end{align}
since the $n=0$ term is an irrelevant constant, and $n=1$ term vanishes upon the integration by parts.
  The appropriate expression for the integral over bosonic  fields $\overline{\psi}\,,\psi$  would differ from Eq.~\eqref{J2}   by the sign in the exponent.

The calculation of the Jacobian in Eq.~\eqref{J2} is rather simple due to the ``loop cancellation theorem" first proved diagrammatically \cite{DzyalLar:73}: only the $n=2$ term contributes in Eq.~\eqref{J2}. For completeness, I give  in Appendix a simple proof of this fact following the proof given in Ref.~\onlinecite{GYL:04}.

The remaining $n=2$ term is straightforward to calculate in the reciprocal space, explicitly using Eq.~\eqref{gf2}, which results in
\begin{align}\notag
  \ln \,& J(\alpha_\eta)=
-\tfrac{1}{2}\Tr(\alpha _\eta g_\eta)^2\\&\notag=
-\tfrac{1}{2}\int\alpha _\eta({\xi })g_\eta({\xi -\xi '})\alpha _\eta({\xi ' })g_\eta({\xi' -\xi })\dd\xi \dd\xi '
\\
\notag&\mapsto
-
\tfrac{1}{2}\int\!
 \alpha _\eta ({\bm k}) \alpha _\eta ({-\bm k}) \tfrac{ \dd^2\bm k }{(2\pi )^2}
\int\!
g_\eta({\bm p})\,g_\eta(\bm{ p{+}k}) \tfrac{ \dd^2\bm p }{({2\pi })^2}\\
&=
\tfrac{\im \eta}{4\pi }\int\!
 \alpha _\eta ({\bm k}) \alpha _\eta ({-\bm k})
{k}g_\eta({\b k}) \tfrac{ \dd^2\bm k }{(2\pi )^2}\notag\\
&\mapsto
\frac{\eta}{4\pi }\Tr\LR \{{\alpha _\eta\,\partial _x\LR [{g_\eta \alpha _\eta }]}\},
\label{J3}\end{align}
where we have used the notations $\bm k\equiv k,\,\omega$ and $\bm p\equiv p,\,\varepsilon $, performed the pole integration over the Green's functions given by Eq.~\eqref{gf2}, and implicitly returned to the $x$-$t$ representation in the last line. Now, using the expressions $\alpha _\eta \equiv  \partial _\eta\theta_\eta$ and  $g_\eta =-\im \partial _\eta^{-1} $, we see\footnote{%
%This is THE FOOTNOTE
One can straightforwardly calculate $J(\alpha_\eta)$ without such a formal manipulation by explicitly writing the last line of Eq.~\eqref{J3} in the $x$-$t$ representation, integrating by parts and using the definitive Green's function equation.
}  that $g_\eta\alpha _\eta=-\im \theta_\eta$.  Thus the full Jacobian, i.e.\ the product of those corresponding to the right- and left-movers is reduced to
\begin{align}\label{J-fin}
    J({\theta_+,\theta_-})=J_+J_- &=\exp\LR [{-\!\!\sum_{\eta=\pm} \frac{\im\eta}{4\pi }\Tr\partial _\eta\theta_\eta\,\partial _x\theta_\eta}]\!\! .
    \end{align}

  Similar considerations work for calculating the Jacobian of the bosonisation transformation, Eq.~\eqref{change}. For the given chiral component, $\eta=\pm$, we have
  \begin{align*}
    \int\mathcal{D}\overline{\psi}_\eta\mathcal{D}
\psi _\eta\,\e ^{\im {   \mathcal{S}}_{0\eta} } =\int{\mathcal{D}}\theta_\eta J ({\alpha _\eta})\int\mathcal{D}\chi _\eta \e ^{\im \widetilde{   \mathcal{S}}_{0\eta} }
  \end{align*}
Since ${   \mathcal{S}}_{0\eta}$ vanishes, Eq.~\eqref{L}, the integral over Majorana fields is reduced to a constant that vanishes in calculating any correlation function. It plays the same role as the Klein factor in the standard operator bosonisation  \cite{vDSh:98,Giamarchi} and is relevant, in the absence of single-fermion excitations, only when there are more than one channel of the same chirality. Then only the integral over the Jacobian is remaining so that the entire contribution to the free action of the $\eta$ (i.e.\ the right or left) -mover is
\begin{align}\label{S0eta}
\mathcal{S}_{0\eta}=-\im\ln J(\alpha_\eta)=-\frac{\eta}{4\pi }\Tr\partial _\eta\theta_\eta\,\partial _x\theta_\eta
\end{align}

Using $\partial _\pm\equiv \partial _t\pm \vF\partial _x$, one finds the full free action, $\mathcal{S}_0=\mathcal{S}_{0+}+\mathcal{S}_{0-} $, in terms of the new bosonic variables,
\begin{align}\label{tf}
    \theta&=\tfrac{1}{2}\LR ({\theta_{\mathrm{L}}-\theta_{\mathrm{R}}}),&\varphi &=\tfrac{1}{2}\LR ({\theta_{\mathrm{L}}+\theta_{\mathrm{R}}}),
\end{align}
as follows:
\begin{align}\label{S0}
  \!\!  \mathcal S_0\!=\!\int\!
\LR \{{\partial _t\theta\,\partial _x\varphi +\partial _t\varphi \,\partial _x\theta- \vF[\LR
 ({\partial _x \theta})^2-  \LR
 ({\partial _x \varphi })^2]}\}\!\tfrac{\dd\xi}{2\pi }.
\end{align}
By integrating by parts, the first two terms can be written together either as   $2\partial _t\theta\, \partial _x\varphi  $      or $2\partial _t\varphi \, \partial _x\theta$, so that   the integrand in   Eq.~\eqref{S0} has the structure like        $\mathcal{L}=\dot qp-\mathcal{H}$                         implying that either $\theta$ or $\varphi $ can be considered as a generalised coordinate, with          $\partial _x\varphi $ or $\partial _x\theta$          being the corresponding generalised momentum. The corresponding Hamiltonian is then
\begin{align*}
    \mathcal{H}_0&=\frac{v}{2\pi }\int\LR [{\LR ({\partial _x\theta})^2+ \LR ({\partial _x\varphi })^2}],
\end{align*}
which is equivalent to Eq.~\eqref{H0}. For what follows, it is more convenient to deal with the action written either in terms of $\theta$, or in terms of $\varphi $. As Eq.~\eqref{S0} is quadratic in both these variables, integrating either of them we end up with the dual representation of the quadratic action:
\begin{subequations} {
\begin{align}\label{S0theta}
     \mathcal{S} _0&=\int\!
\LR [{\LR ({\partial _t \theta})^2-\vF^2\LR
 ({\partial _x \theta})^2}]\tfrac{\dd\xi }{2\pi \vF}\\&=
\int\!
\LR [{\LR ({\partial _t \varphi})^2-\vF^2\LR
 ({\partial _x \varphi})^2}]\tfrac{\dd\xi }{2\pi \vF}\
.
\label{S0phi}
\end{align}
}\label{S0d}\end{subequations}
The great advantage of the bosonised action of Eq.~\eqref{S0d} over the original one, Eq.~\eqref{S}, is that the density-density or current-current interaction, which is quartic in terms of the original fields, remains quadratic in terms of ``slow" fields $\theta$ or $\varphi $. The reason is that these fields, introduced by Eq.~\eqref{tf}, represent density and current fluctuations, respectively.

In order to demonstrate this, note that the na\"\i{ve} substitution of the bosonised fields \eqref{change} into the density field $\rho_\eta\equiv \overline{\psi}_\eta\psi_\eta$ leads to the vanishing result as in Eq.~\eqref{L} due to the very same field-theoretical anomaly. To find out how bosonisation Eq.~\eqref{change} works for the density field, one  represents its quantum-mechanical average with the help of a source field $h$ as
\begin{align}
\notag
\lr<>{\rho_\eta}&=\frac{1}{Z}\int\!
\mathcal{D}\overline{\psi}_\eta \mathcal{D}\psi_\eta\LR ({
\overline{\psi}_\eta \psi_\eta\,
\e ^{\im S_{\eta0}} })
\\&=\fdv{}{h_\eta} \eval{ {\int\!\mathcal{D}\overline{\psi}_\eta
\mathcal{D}\psi\e ^{i\int\!\overline{\psi}_\eta({\partial _\eta}-\im h_\eta)
\psi_\eta\,} } }_{h\to0}\label{rho},  \end{align}
and evaluates the Jacobian of substitution \eqref{change} taking the source field into account. This gives, by a full analogy with the derivation of Eq.~\eqref{J}, the following formal expression:
\begin{align}\label{Jh}
    J\LR ({\alpha _\eta;h_\eta})= \frac{\det (\im \partial _\eta-\im h_\eta)}{\det\LR ({\im \partial _\eta-\alpha _\eta -\im h_\eta})}=\frac{J({\alpha _\eta+\im h_\eta})}{J({\im h_\eta})},
\end{align}
where the numerator and denominator of the last fraction are obtained from Eq.~\eqref{J3} by substituting $\alpha _\eta\to\alpha _\eta+\im h_\eta$ and $\alpha _\eta\to \im h_\eta$, respectively. This immediately gives the generating function:
\begin{align}\label{S0h}
    S_{0\eta}\LR ({h_\eta})=S_{0\eta}-\frac{\im \eta}{2\pi } \int\!h_\eta \,\partial _x\theta_\eta\dd\xi \,.
\end{align}
Using this for calculating $\langle{\rho_\nu}\rangle$ in Eq.~\eqref{rho} after  substitution \eqref{change} allows one to express the density field of the $\eta$-movers, as well as the full density,  is  in terms of bosonic variables. The full density is then given by \begin{subequations} {
\begin{align}
 \rho_\nu&=\frac{\eta}{2\pi } \partial _x\theta_\eta\label{rhoeta}\\ \label{fullrho}
    \rho&=\rho_++\rho_-=\frac{1}{2\pi }\partial _x\LR ({\theta_{\mathrm{R}}-\theta_{\mathrm{L}}}) \equiv -\frac{1}{\pi }\partial_x\theta
\end{align}}\label{rho2} \end{subequations}
This relations can be used for easily incorporating into the action any density-density interaction. A similar expression for the current in terms of the dual field, $j=\frac{1}{\pi }\partial _x\varphi $ can be used for incorporating the current-current interaction.

Allowing for the fact that interaction coupling constants for movers of opposite or the same chirality might be different, a  Hamiltonian for the short-range (contact) density-density interaction  can be represented (using the traditional notations for the coupling constants) as
\begin{align}\label{H_int}
    \mathcal{H}_{\mathrm{int}}&=\int\!\Big [{\tfrac{1}{2}g_4\!\sum_{\eta={\mathrm{R,L}}} \rho_\eta^2+g_2\rho_{\mathrm{R}}\rho_{\mathrm{L}}
    }\Big]\dd x.
\end{align}
Using Eq.~\eqref{rhoeta}, this can be rewritten as a quadratic in $\partial _x\theta_\eta$ expression similar to that for the free particles Hamiltonian of Eq.~\eqref{H0}. Adding these expressions, one obtains the full Hamiltonian of the Luttinger liquid with density-density interaction as
\begin{align*}
     \mathcal{H}&=\mathcal{H}_0+\mathcal{H}_{\mathrm{int}}\equiv \int\!H({x})\,\dd x
\end{align*}
where the Hamiltonian density is found from Eqs.~\eqref{H0}, \eqref{rho2} and \eqref{H_int} as
\begin{align}\notag
   H&({x})
    =\frac{\pi \vF {+}\frac{1}{2}g_4}{4\pi ^2}\sum_{\eta={\mathrm{R,L}}} \LR ({\partial _x\theta_\eta})^2-\frac{g_2}{4\pi ^2}{\LR ({\partial _x \theta_{\mathrm{R}}})}\LR ({\partial _x \theta_{\mathrm{L}}})\notag\\\notag&=\frac{\pi \vF {+}\frac{1}{2}g_4}{2\pi ^2}\!\LR [{\!\LR ({\partial _x\theta})^2 {+}\LR ({\partial _x\varphi })^2}]-\frac{g_2}{4\pi ^2}\!\LR [{\!\LR ({\partial _x\varphi })^2{-}\LR ({\partial _x\theta})^2 }]
    \\&\equiv \frac{v}{2\pi K}\LR ({\partial _x \theta})^2+ \frac{vK}{2\pi }\LR ({\partial _x \varphi })^2\label{H},
\end{align}
where the Luttinger parameter $K$ and the effective velocity $v$ are defined by
\begin{align}\label{v+K}
K&=\sqrt{\frac{2\pi \vF+g_4-g_2}{2\pi \vF+g_4+g_2}}\,,
    &v&\equiv \frac{\vF}{K}+\frac{g_4-g_2}{2\pi K}\,.
\end{align}
In many cases (e.g., the screened Coulomb interaction), one has $g_2=g_4\equiv g$ so that expressions \eqref{v+K} are simplified to $K=\LR ({1+g/2\pi \vF})^{-1} $ and $v=\vF/K$. Note that $K<1$ for the repulsive interaction (when $g_2>0$), while  $K=1$ corresponds to the free particle case since Eq.~\eqref{H} coincides in this case with Eq.~\eqref{H0}.

More general interactions between fermions can still be encompass with Hamiltonian \eqref{H} where positive $K<1$ and the effective Fermi velocity $v$ should be considered as the phenomenological parameters.\cite{Giamarchi}
The action corresponding to Hamiltonian \eqref{H} has a dual representation in terms of either the density field $\theta$, or the current field $\varphi $ similar to that of Eq.~\eqref{S0d}:
\begin{subequations} {
\begin{align}\label{Stheta}
     \mathcal{S} _0&=\frac1{2\pi vK}\int\!
\LR [{\LR ({\partial _t \theta})^2-v^2\LR
 ({\partial _x \theta})^2}] {\dd\xi }\\[3pt]&=\frac K{2\pi v}
\int\!
\LR [{\LR ({\partial _t \varphi})^2-v^2\LR
 ({\partial _x \varphi})^2}]\dd\xi
.
\label{Sphi}
\end{align}
}\label{Sd}\end{subequations}

One arrives at exactly the same action when describing interacting bosons above the quasi-condensate in 1D. In this case, one starts with Bogolyubov's solution of the Gross--Pitaevskii equation (see, e.g., \cite{Pethick-Smith}), $\varepsilon _k=\sqrt{\varepsilon _{k0}
\LR ({\varepsilon _{k0} +2gn})}
$, which is naturally linearised for $\varepsilon _{k0}\equiv k^2/2m\ll gn$: $\varepsilon _k\to ck$ where $g$ is the interaction constant, $n$ is the average 1D particle density and  $c=\sqrt{gn/m}$ is the speed of sound. I will not go into details of \emph{bosonisation} (i.e.\ switching to the density-phase variables) in this case, as techniques are similar. It is sufficient to note that the left- and right-movers are introduced via the exponentiation of the (already bosonic) field $\psi({x, t})$ as  \begin{align}\label{bosbos}
\psi (x,t) =\sqrt{n+\partial _x\theta(x,t)\,}\, \mathrm{e} ^{i\varphi(x,t)},
\end{align}
where similarly to the fermionic case $\partial _x\theta(x,t)$ describes the density fluctuations, and $\varphi ({x,t})$ is the phase field. The resulting action coincides with Eq.~\eqref{Sd} but with \mbox{$v\to c$}, and with the Luttinger parameter $K$ related to the average density $n$ and the healing length $\xi =1/mc$ by $K=\pi n\xi $. For a typical cold-atomic gas, the healing length is much greater than the average inter-particle distance so that $K\gg1$.

\section{1D Bosonic flow through tunnelling junction}
Now I illustrate how to use the developed techniques for describing the  bosonic flow between BEC reservoirs connected by a 1D channel through weak links,\cite{SGKL} see Fig.\ref{fig1}. The corresponding action is $\mathcal{S}=\mathcal{S}_0 +\mathcal{S}_J$, where $\mathcal{S}_0$ is given by Eq.~\eqref{Sphi} (the phase representation is, obviously, preferable), and $\mathcal{S}_J$ describing the tunnelling through the weak links is given by %
\begin{align}\label{ST}
S_\mathrm{T}&=2J \int\!\mathrm{d}\xi
  \big[ \cos\phi_\mathrm{R}\delta(x{-}\tfrac{L}{2} ) +
  \cos\phi_\mathrm{L}\delta(x{+}\tfrac{L}{2})\big],
\end{align}
where $\phi _{\mathrm{L,R}}$ are the expected phase jumps at the tunnelling barriers. Note the difference in notations between the boundary phase fields $\phi _{ {\mathrm{L,R}}} ({t})$ and the `bulk' phase fields $\varphi ({x,t})$. These fields are related by the boundary conditions
\begin{align}\label{phi}
    \phi_\mathrm{L} &=\Phi -\varphi(- L/2)\,, &\phi _\mathrm{R}&=\Phi +\varphi  (L/2)\,.
\end{align}
We assume that the left and right reservoirs have many more particles than in the 1D channel and are maintained at the fixed phases equal $\pm \Phi $. We have chosen, without loss of generality, the tunnelling energies $J$ being equal at both the barriers (only a trivial rescaling is required if they are not).  As usual,  the tunnelling action is valid when the overlap of the wave functions across the barrier is small,  which imposes  the requirement $J\ll cK/\xi\equiv \pi nc$.

On the face of it, this configuration is very similar to the electronic one: two usual superconductors connected via weak tunnelling links by a 1D wire. \cite{Fazio,*Fazio1,*MasStoneGoldbartLoss,*AffleckCauxZ} When the contacts are ideal rather than tunnelling, the interaction makes no impact so that the critical superconducting current does not depend on the interaction parameter $K$:
\begin{align*}
{\mathcal I}_c=
\mathcal{I}_{\mathrm{c}0}
\frac{\xi\Delta \Phi }{L},%{J^2 \Lr(){\frac L\xi}{1/K_\rho+1/K_\sigma-2}}
\end{align*}
where $\mathcal{I}_{\mathrm{c}0}$ is a critical current in a bulk superconductor, and $L$ is the length of the wire.
This is similar to the case of conductance of a normal 1D wire which -- in the absence of impurities -- is not affected by the interaction. \cite{MasStone:95} This changes in the case of the weak links where the interaction leads to the further suppression of the critical proximity superconducting current:
\begin{align*}
{\mathcal I}_c=
\mathcal{I}_{\mathrm{c}0}
\frac{\xi\Delta \chi}{L}{J^2 \lr(){\frac \xi L}^{1/K-1}}.
\end{align*}
Obviously, this cannot be generalised to the case of repulsive bosons where $K>1$ (or $\gg1$ in a typical for cold atoms case), and the power of the small parameter becomes negative. This means that the perturbative (in the powers of $J^2$) approach fails, and the starting point must be different.

\begin{figure}
\centering
\includegraphics[width=.95\columnwidth]{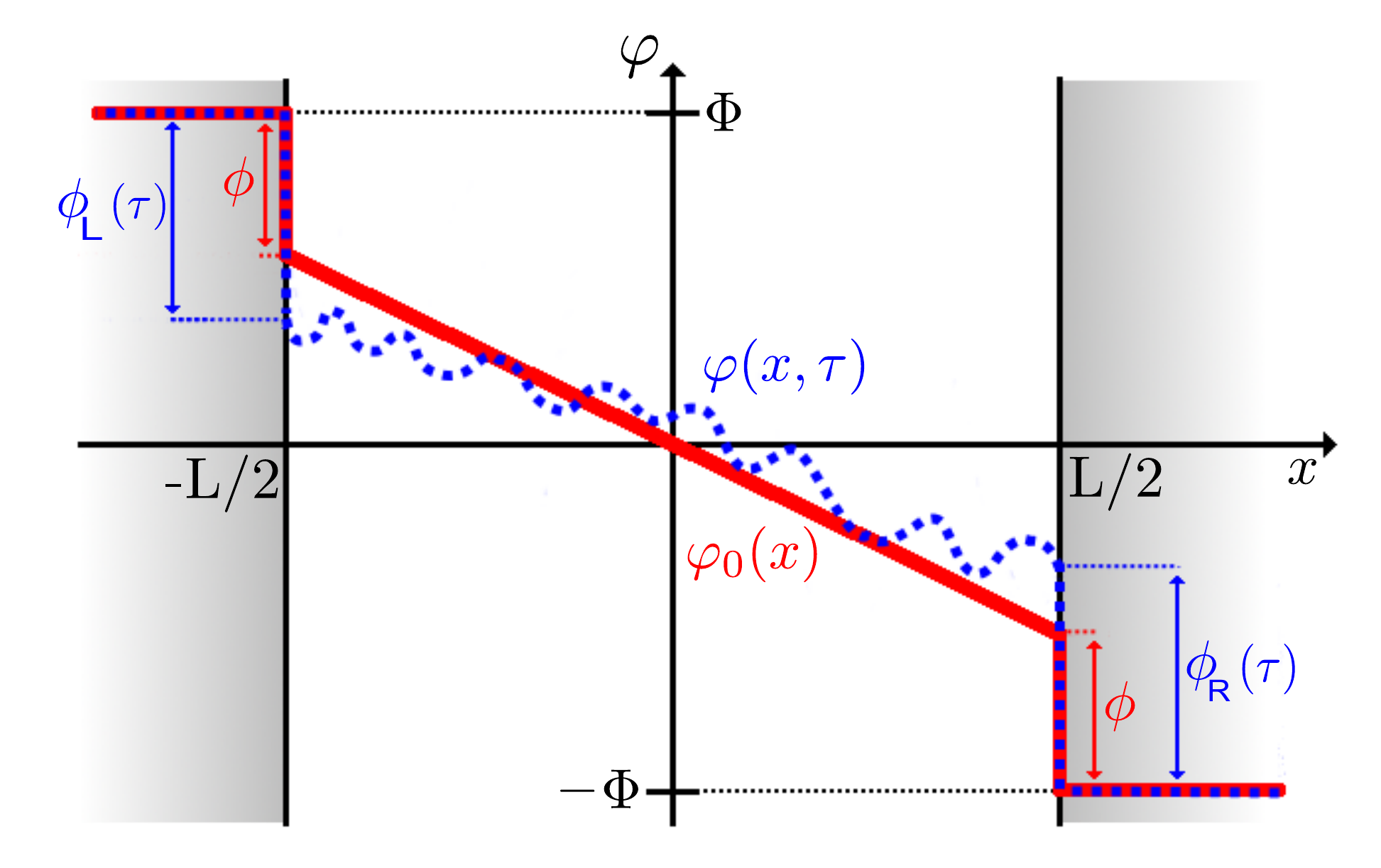}
\caption{(Color online) Phase profile along the
  channel. The solid line, $\varphi  _0({x})$ is a typical (symmetric) configuration made up of a linear superfluid contribution and phase
  jumps, $\phi  $, at  each tunnel barrier.  The dashed line represents a fluctuation around the phase profile.  Here we have chosen the phases of the BEC reservoirs as $\Phi _\mathrm{L}=-\Phi _\mathrm{R}\equiv \Phi   $. The phase profile shown above corresponds to  a phase difference of $2\Phi <\pi $.}\label{phasefig}
\end{figure}

Such a starting point is finding a stationary (i.e.\ $t$-independent) mean-field (MF) configuration for  the model described by the action of Eqs.~\eqref{Sphi} and \eqref{ST}. Since the tunnelling part  \eqref{ST} is $x$-independent, and the bulk part \eqref{Sphi} contains the $2^{\mathrm{nd}}$ derivative, the action is optimised by the linear in $x$ phase field $\varphi _0({x})$ that satisfies the boundary conditions \eqref{phi}:
\begin{align}\label{condition}
   \varphi_0(x)&=2({\phi _+-\Phi  })\frac{x}{L}-\phi _-\,,  & \phi_\mathrm{\pm}  &\equiv\frac{1}{2}({\phi _{\mathrm{L}}\pm\phi _\mathrm{R} })\,.
\end{align}
This phase profile is illustrated in Fig.~\ref{phasefig}.
It describes a constant superflow, $\mathcal{I}= nv $,  between
the reservoirs, with a velocity $v=-2({\Phi -\phi _+})/mL$.
The energy $E$ is the sum of  the supercurrent kinetic energy,  $\frac12{mN}v^2$, which arises from the Luttinger action (\ref{Sphi}) on substituting ansatz ({\ref{condition}}),  and the Josephson energy,
$
  - 2J(\cos\phi _\mathrm{R}+\cos\phi _\mathrm{L})\,.
$
The total  dimensionless energy, $\varepsilon  \equiv E/J_\mathrm{c} $,  can be written via the phase jumps $\phi _\pm$ as
\begin{align}\label{eq:energy}
   \varepsilon  &=  \phantom{-}2({\Phi -\phi_+  })^2 -4\alpha \cos \phi_+\cos\phi _-  \,, &\alpha& \equiv  J/J_\mathrm{c}
\end{align}
where $J_\mathrm{c}\equiv n/mL \ll \pi  nc$  so that $\alpha $ can vary from $0$ to values $\gg1$ within the region of applicability of the tunnelling action, Eq.~(\ref{ST}).

The next step is to minimise the energy with respect to the phase jumps $\phi _\pm$ for the fixed phase difference $2\Phi $ between the reservoirs. The minimisation conditions are \begin{subequations}\label{MF}\begin{align}
  \label{eq:phiphi}
 &\Phi -\phi _+ = \alpha \sin\phi_+\cos\phi _-\,,\\
   &\cos\phi _+\sin\phi _- =0\,.\label{phi-}
\end{align}
Since energy ({\ref{eq:energy}}) is a $2\pi $ periodic function of $\phi  _-$, we can restrict ourselves to  two solutions of Eq.~(\ref{phi-}), corresponding to  the symmetric phase drops, $\phi  _-=0$ so that $\phi  _\mathrm{R}=\phi  _\mathrm{L}  =\phi  _+$, and asymmetric ones, $\phi  _-=\pi $ so that  $\phi  _\mathrm{L}  =\phi  _++\pi $. Solutions corresponding to $\cos\phi  _+=0$ are always unstable (saddle points). For the  symmetric/asymmetric branch Eq.~({\ref{eq:phiphi}}) is  reduced to
   \begin{align}\label{pm}
   \Phi -\phi  _+=\pm \alpha \sin\phi  _+
\end{align}\end{subequations}
 The symmetric-branch equation is almost identical to  that emerging in a text-book analysis  \cite{Tinkham} of a SQUID; however, its solution has a peculiar $4\pi $ periodicity. It is the coexistence of this solution with that for the asymmetric branch which restores the correct $2\pi $ periodicity.  Indeed, each of Eqs.~(\ref{pm}) has at least one stable solution in some interval of $\Phi $ and  these intervals always overlap.

\begin{figure}
\centering
\includegraphics[width=0.9\columnwidth]{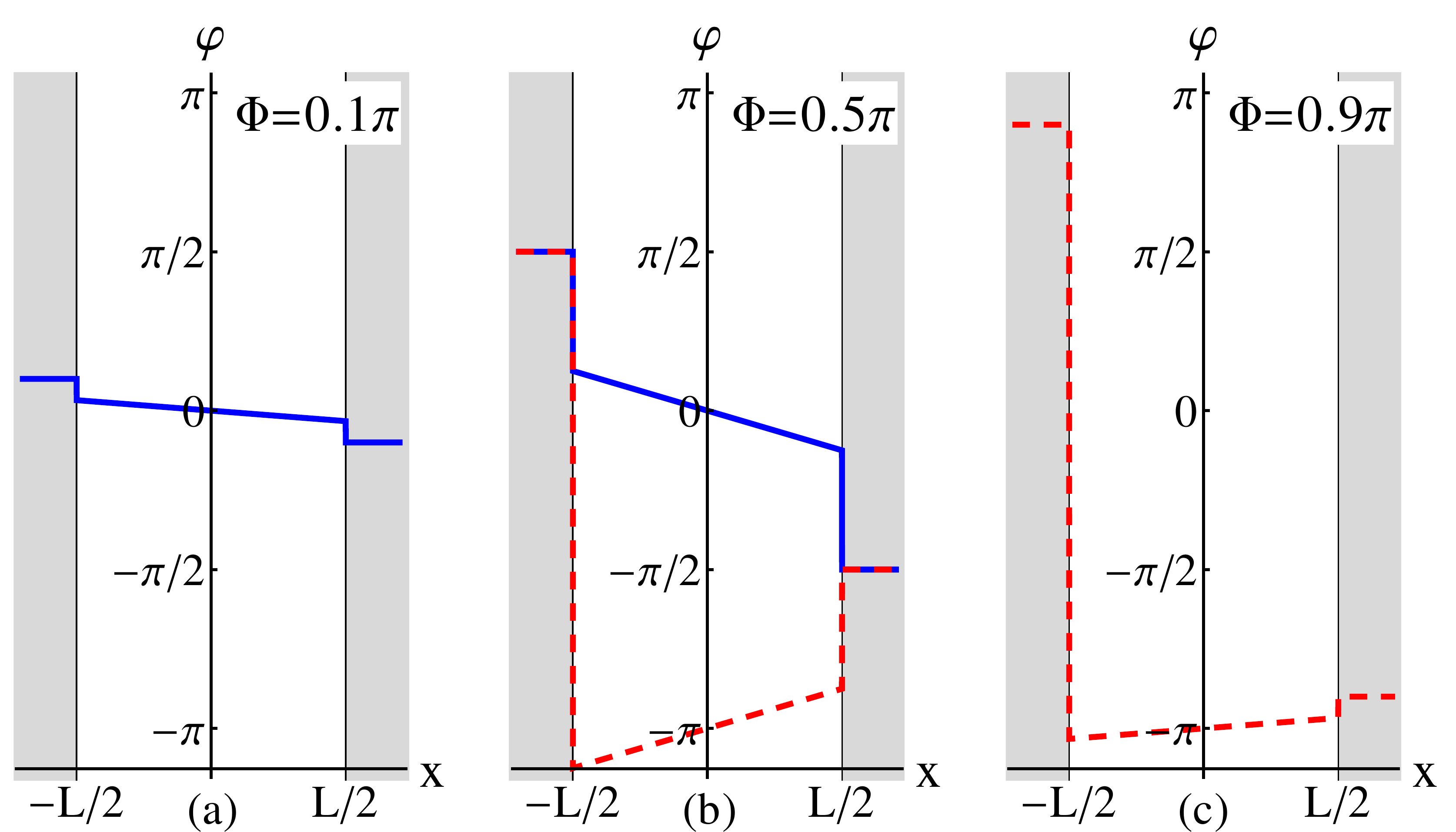}
    \caption{(Color online)  The MF phase profile in the channel for $\alpha\!<\!1$ ($J\!<\!J_\mathrm{c} $): (a) and (c) are unique symmetric/asymmetric solutions near $\Phi\!=\!0 $ or $\pi $, respectively; (b)~these two solutions become degenerate at $\Phi =\pi /2$,   with one of them  becoming metastable slightly above or below $\pi/2$.\label{MultiSol} }
\end{figure}

\begin{figure}[t!]
\centering
\includegraphics[width=0.95\columnwidth]{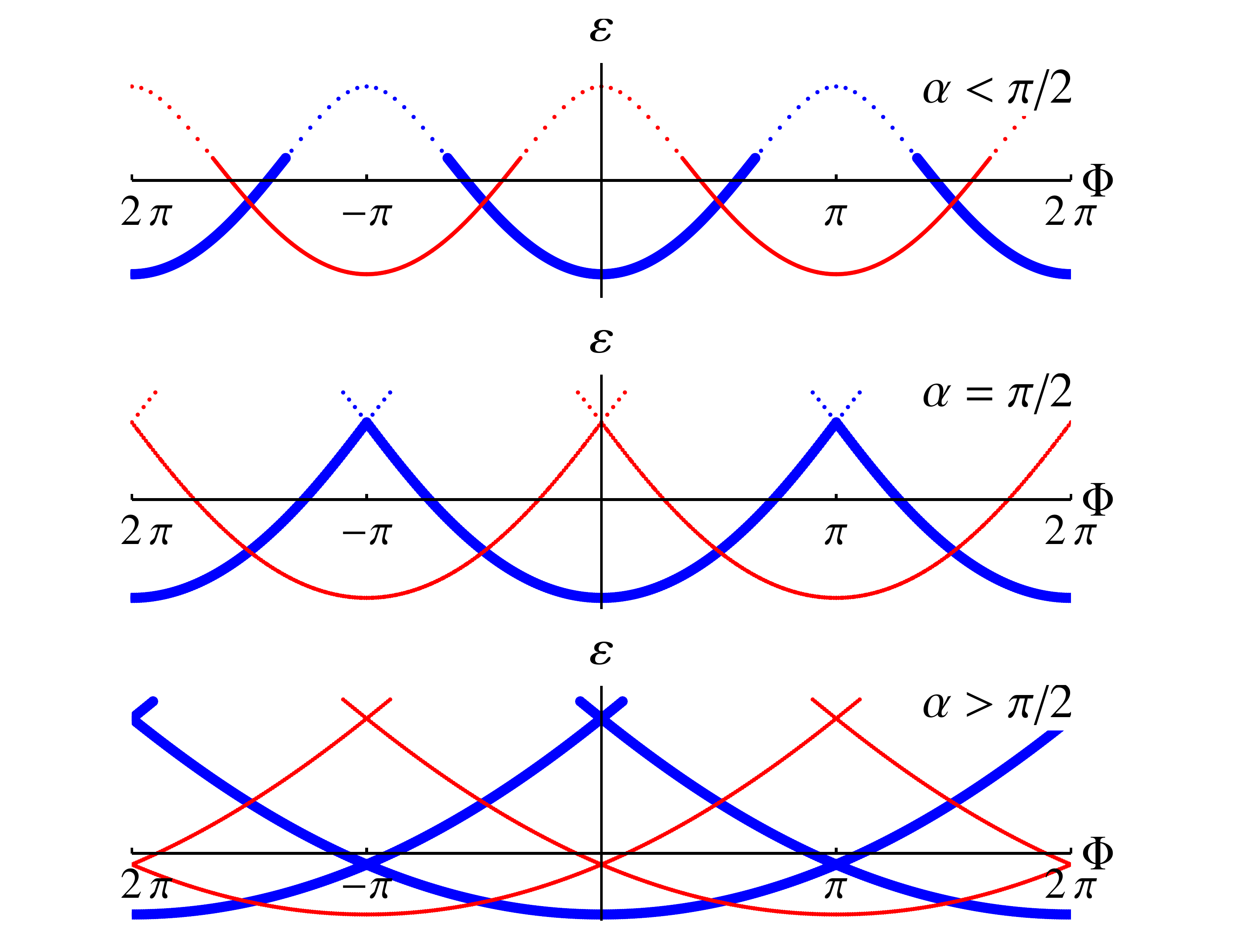}
(a)

\vspace*{12pt}

\caption{(Color online)  The MF energies ({\ref{eq:energy}}), spread between $\varepsilon  _\mathrm{min}= -4\alpha $ and $ \varepsilon_{\mathrm{max} }=2\alpha ^2$,  as functions of the external phase difference, $2\Phi $, at different values of $\alpha$. Thick (thin) solid lines represent symmetric (asymmetric) stable or metastable   solutions,   the latter lying in the continuum of phononic excitations.
}\label{cusps}
\end{figure}

The MF energy   is thus no longer a single-valued function of $\Phi $. Assuming first a singly connected geometry, when the external phase difference $2\Phi \in[{0,2\pi }] $,
we find for small $\Phi $ that the lowest energy solution of Eq.~(\ref{eq:phiphi}), which is $\phi _+\approx \Phi/(1+ \alpha) $, belongs to the symmetric branch. An elementary analysis shows that for small $\alpha $ it remains stable with increasing $\Phi $  up to $\Phi =\pi /2+\alpha $. The lowest-energy solution around $\Phi =\pi $ belongs to the asymmetric branch and remains stable down to $\Phi =\pi /2-\alpha $. Thus, in the interval of width $2\alpha $ centred at \mbox{$\Phi =\pi /2$}  the two solutions   coexist: the symmetric solution is stable and the asymmetric is metastable at \mbox{$\Phi <\pi /2$}, with their roles reversing at $\Phi >\pi /2$, as illustrated in Fig.~{\ref{MultiSol}}.
With $\alpha $ increasing, two new solutions appear at $\alpha >1$ for both the symmetric and asymmetric branch but they remain unstable until $ \alpha $ reaches $\pi /2$. At this point the two solutions coexist in the entire   interval $[{0,\pi }] $, while new metastable solutions emerge for the asymmetric branch around $\Phi =0$ and for the symmetric around $\Phi =\pi $. With $\alpha $ further increasing, new pairs of metastable solutions appear at integer multiples of $\pi /2$, see Fig.~{\ref{cusps}}. There are multiple crossing of different MF branches; the lowest crossing correspond to double degeneracy of the ground state at the phase difference $2\Phi =\pi +2\pi n$.

One would expect that quantum fluctuations due to phonon excitations above the quasi-condensate would lift this degeneracy and result in avoided crossings.  However, this turned out not to be the case\cite{SGKL} as I will demonstrate in the following section.

\section{Quantum fluctuations around the MF solution}\label{fluct}
\begin{figure}[b]
\begin{center}
\includegraphics[width=\columnwidth]{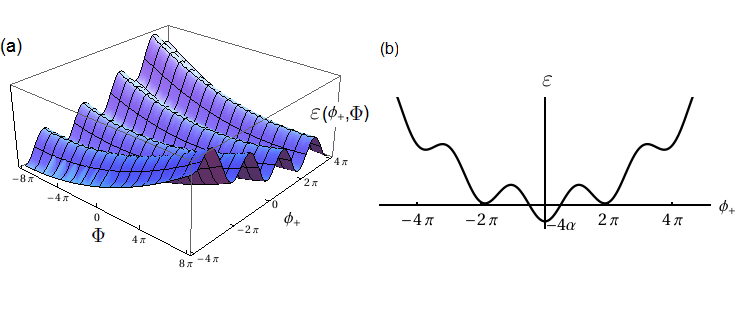}
\end{center}
\caption{\label{washb} (color online)  (a) A sketch of the energy dependence on the phase difference between the reservoirs, $2\Phi $, and the MF value $\phi _+$ in the symmetric case, $\phi _-=0$; (b) $\varepsilon ({\phi _+})$-dependence at $\Phi =0$.}
\end{figure}

It is remarkable that quantum fluctuations do not result in avoided crossings in Fig.~\ref{cusps}(a). To show this we introduce the phase fluctuations in the 1D channel, $\widetilde\varphi(x, t )= \varphi(x, t )-\varphi_0(x)$ and at the boundaries, $\widetilde\phi  _{\mathrm{L,R} }( t )= \phi  _{\mathrm{L,R} }( t )-\phi_{\mathrm{L,R} }$, related by the boundary conditions $\widetilde\varphi(\pm L/2, t ) = \pm \dphi_{\mathrm{L,R} }( t )$. Here
$\varphi  _0({x})$ and $\phi_{\mathrm{L,R} } $ are the  solutions of the MF equations ({\ref{MF}}) described above, related to the symmetric and antisymmetric combinations  introduced in Eq.~(\ref{condition}). Since the action is quadratic in the 1D channel, with nontrivial parts being only at the boundary, we obtain the effective $0+1$ action after integrating out the bulk fields $\widetilde{\varphi} $:
\begin{align*}
 S_\mathrm{eff}  &= K\!\int \! \frac{\mathrm{d}\omega}{2\pi^2} \left[  \omega \coth\left(\frac{\omega}{\omega_0}\right) |\dphi_{+}(\omega)|^2 +  \frac{\omega^2}{\omega_0} |\dphi_{-}(\omega)|^2 \right]\nonumber \\ &+\int \! \mathrm{d} t   \Big[  2nv \dphi_{+}( t ) +4J\cos\dphi_{-}( t ) \cos( \phi \!+\! \dphi_{+}( t ))\Big],
 \end{align*}
At relatively low energies, $\omega\ll c/L$ (corresponding to the times sufficient for multiple traversing the channel), the action is simplified to the Caldeira-Legget form,
 $S_{\mathrm{eff}}=S_\mathrm{fl}+S_\varepsilon   $, with
\begin{subequations}\label{effectiveaction}
\begin{align}\label{Ohm}
 S_\mathrm{fl}
 &= K\!\int \! \frac{|\omega |\mathrm{d}\omega}{2\pi^2} \left[    |\dphi _+(\omega)|^2 +    |\dphi_  -(\omega)|^2 \right],
 \\\label{eps}
  S_\varepsilon  &
 =\!\int \!\! \mathrm{d} t   \Big[
2({\Phi -\widetilde\phi_+  })^2 -4\alpha \cos \widetilde\phi_
+\cos\widetilde\phi _-  \Big]
\end{align}
\end{subequations}
Here $S_\varepsilon  $   plays the role of an effective ``washboard'' potential for the Caldeira-Legget type   action of Eq.~(\ref{Ohm}), see Fig.~\ref{washb}.

Now we perform the standard renormalization group  (RG) analysis by integrating out fast modes in the fields $\phi  _+ $ and $\phi  _- $ in the standard way.\cite{KaneFis:92a} This  results in the RG equation for the dimensionless tunnelling strength $\alpha   $:
\begin{align}\label{RG}
\frac{\dd \ln\alpha }{\dd \ln b}=1- \frac{1}{2K},
\end{align}
where b is a scaling parameter.  The integration between the upper, $\Lambda \sim c/\xi $, and lower, {$\omega _0\sim \max\{{T,\, c/L}\}  $}, energy cutoffs  gives the renormalized dimensionless tunnelling as $\alpha ({\omega _0 }) = \alpha _0 \left( \Lambda /\omega _0 \right)^{1-\frac{1}{2K}}$, where $\alpha _0\equiv J/J_\mathrm{c} $. Since the tunnelling through barriers separated by $L\gg\xi$ is uncorrelated, this is similar to the results   for  tunnelling through a single barrier \cite{KaneFis:92a}, as well as to the results for superconducting systems \cite{Fazio} in a geometry similar to that under consideration here. However, the consequences are very different. For superconducting systems with $2K<1$, the tunnelling strength $\alpha $ flows to zero, so that this parameter is irrelevant in the RG sense. Thus, the MF solution is washed out and the perturbative in tunnelling approach\cite{Fazio} is fully justified.

On the contrary, for $K>1$ (actually,  $K\gg1$  for ultracold bosonic systems with the healing length much bigger than the interatomic distance) $\alpha $ flows to larger values. This means that the washboard potential  becomes more pronounced so that the fluctuations are  irrelevant in the low-energy limit and the MF solution, described above, is robust. In particular, since the fluctuations do not connect different MF branches, the level crossings are not avoided and the characteristic cusps in energy, Fig.~\ref{cusps}, and the corresponding jumps in the superflow remain.

\begin{figure}[t]
  \includegraphics[width=.4\textwidth]{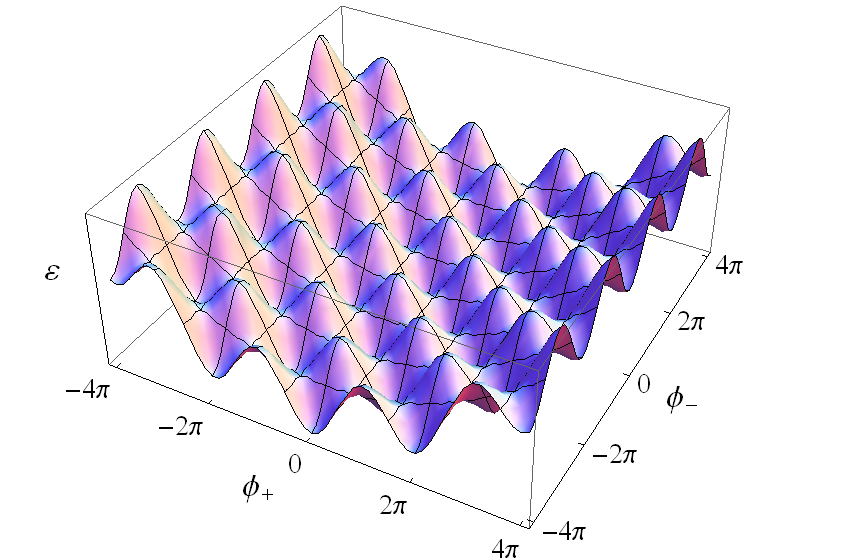}\\
  \caption{A graph showing the energy profile $\varepsilon ({\phi _+, \,\phi _-})$. The instanton moves
from one minimum of this profile to a neighbouring minimum.}\label{eggs}
\end{figure}

Alternatively, this can be seen using instanton techniques similar to those of Ref.~\onlinecite{Buchler2001,*Schecter2012}. An instanton is a saddle-point solution of a classical (imaginary time) action giving the probability of tunnelling between adjacent minima. In the present case, the adjacent minima always correspond to symmetric, $\phi _-=0$, and asymmetric, $\phi _-=\pi $, solutions, see Fig.~\ref{eggs}.
Making the Fourier transform to the imaginary time representation in the action \eqref{Ohm} results in \begin{align} \notag
S_{\mathrm{fl}} =  -\frac{K}{\pi^2}  \int \mathrm{d}\tau_1 \mathrm{d}\tau_2 \, \sum_{\eta=\pm}\dot{\phi}_\eta(\tau_1) \log\left\lvert\frac{(\tau_1 - \tau_2)}{\tau_{\mathrm{c}}}\right\rvert \dot{\phi}_\eta(\tau_2) ,
 \end{align}
which should be treated together with the tunnelling action \eqref{eps} in the imaginary-time representation. In the lowest order, the instanton configuration connecting the nearest minima can be approximated for both fields as $\dot\phi _\pm=\pi [{\delta ({\tau}) -\delta ({\tau-\overline{\tau}})}]$, where $\overline{\tau}$ is  the time spent in the metastable state. Substituting these configurations
into the above action results in the effective single-instanton action
  \begin{align}\label{inst}
S_{\mathrm{inst}} = 4K \log \left\lvert \frac{\bar{\tau}}{\tau_{\mathrm{c}}}
 \right\rvert -  2\pi\bar{\tau}  J_{\mathrm{c}} \left[ \pi - 2 (\Phi - \phi_+^0) \right],
\end{align}
which is valid in the vicinity of the crossing point of the two solutions, where the second term above is $\ll1$. Now, making the Wick rotation back to the real time one finds the tunnelling rate between the two close-to-minima configurations as $\Gamma\propto\int\dd t\e^{iS_{\mathrm{inst}}[{it}]}\sim \varepsilon ^{4K-1}  \ll \varepsilon $ for $K>\frac{1}{2}$ in the full agreement with the above RG results. The meaning of this is that when the energy difference between the two configurations is decreasing, the rate of tunnelling is decreasing much faster, completely vanishing at the exact resonance. Although this is `only' the power-law decrease, it is hard to distinguish it from an exponential suppression at $K\sim 20\divisionsymbol 30$ as in typical cold-atomic gases.

\section{Discussion}
In these lectures, I have shown how to develop the functional bosonisation techniques\cite{GYL:04} and used it for describing   quantum superflow of neutral bosons through a 1D channel connecting two BEC reservoirs with different phases via weak tunnelling links.\cite{SGKL} The superflow has characteristics totally different from those in analogous systems of two superconducting BCS reservoirs connected by a 1D wire.\cite{Fazio} Namely, the superflow is non-perturbative in the tunnelling energy, periodically flips directions with the phase difference changing, and has metastable branches for any tunnelling strength. The energy levels corresponding to different branches intersect, and quantum fluctuations do not lead to avoided crossings. In particular, the ground state remains double-degenerate at the phase difference $\pi $ between the reservoirs, with two intersecting branches both being $4\pi $-periodic.  This resembles the pattern characteristic for the current through a 1D wire connecting two topologically nontrivial superconductors. \cite{Lutchin:2010,*OregRefOpp} The physical reason for such a similarity is that in the latter system backscattering at the edges of 1D channel is topologically forbidden, while in the system described here such a backscattering is dynamically suppressed as described in Section \ref{fluct}. It remains to be seen whether there exists a deeper analogy between the edge excitations, which are dynamical phase jumps in the present case and Majorana fermions in the topologically nontrivial superconductors connected by the 1D wire.\cite{Lutchin:2010,*OregRefOpp}
%\bibliography{my}
%Control: production of eprint (-1) disabled
%

\end{document}